\documentclass[english,nofootinbib,notitlepage]{revtex4}
\pdfoutput=1

\usepackage[T1]{fontenc}
\usepackage[latin9]{inputenc}
\usepackage{color}
\usepackage{graphicx}
\usepackage{amssymb,amsmath}
\usepackage{babel}

\def\be{\begin{equation}}
\def\ee{\end{equation}}

\makeatletter

\@ifundefined{definecolor}
 {\usepackage{color}}{}

\begin{document}

\title{Inflation from the Higgs field false vacuum \\ with hybrid potential}

\author{Isabella Masina$^{1,2}$}
\email{masina@fe.infn.it}
\author{Alessio Notari$^{3}$}
\email{notari@ffn.ub.es}

\affiliation{$^{1}$  Dip.~di Fisica, Universit\`a di Ferrara and INFN Sez.~di Ferrara, Via Saragat 1, I-44100 Ferrara, Italy}
\affiliation{$^{2}$ CP$^3$-Origins \& DIAS, Southern Denmark University, Campusvej 55, DK-5230 Odense M, Denmark}
\affiliation{$^{3}$ Departament de F\'isica Fondamental i Institut de Ci\`encies del Cosmos, Universitat de Barcelona, Mart\'i i Franqu\`es 1, 08028 Barcelona, Spain}

\begin{abstract}
We have recently suggested \cite{Masina:2011aa,Masina:2011un}  that Inflation could have started in a local minimum of the Higgs potential 
at field values of about $10^{15}-10^{17}$ GeV, which exists for a narrow band of values of the top quark and Higgs masses and thus gives rise to a prediction on the Higgs mass to be in the range $123-129$ GeV, together with a prediction on the the top mass and the cosmological tensor-to-scalar ratio $r$. Inflation can be achieved provided there is an additional degree of freedom which allows the transition to a radiation era.
In~\cite{Masina:2011aa} we had proposed such field to be a Brans-Dicke scalar. Here we present an alternative possibility with an additional 
subdominant scalar very weakly coupled to the Higgs, realizing an (inverted) hybrid Inflation scenario. Interestingly, we show that such model has
 an additional constraint $m_H<125.3  \pm 3_{th}$, where $3_{th}$ is the present theoretical uncertainty on the Standard Model RGEs. 
 The tensor-to-scalar ratio has to be within the narrow range $10^{-4}\lesssim r<0.007$, and values of the scalar spectral index compatible with the observed range  can be
  obtained.  Moreover, if we impose the model to have subplanckian field excursion, this selects a narrower range $10^{-4} \lesssim r<0.001$ and an 
  upper bound on the Higgs mass of about $m_H <124 \pm 3_{th}$.
\end{abstract}


\maketitle

\section{Introduction}

In~\cite{Masina:2011aa} we have considered the possibility of realizing Inflation from the Standard Model Higgs sector, 
using the fact that the Higgs potential has a local minimum between $10^{15}-10^{17}$ GeV, which exists for a narrow 
band of the top and Higgs mass values \cite{CERN-TH-2683,hep-ph/0104016,strumia2}.
Not only such a local minimum exists but, within
the allowed parameter range in the Higgs and top masses, its energy density  has the right value (GUT scale)
to give rise to the correct amplitude of density perturbations from Inflation.

In order to get successful exit from Inflation we have embedded the Standard Model in a scalar-tensor theory of gravity.
More precisely we have used two ingredients: 
(i) the existence of a false vacuum in the Higgs potential which can source exponential expansion in the early Universe; this gives rise to a 
combined prediction on the Higgs mass $m_H$, the top quark mass  $m_t$ and the ratio of tensor to scalar perturbations from 
Inflation $r$, all of which can be experimentally checked in the near future (ii) the possibility of achieving a transition from an exponential expansion
to a hot radiation era by adding an extra scalar in the gravitational sector of the theory, which slows down the expansion allowing 
a tunneling of the Higgs field.

In~\cite{Masina:2011un} we have pointed out that the combined predictions on $m_H$, $m_t$ and $r$ are generic 
and independent on the way in which the exit from Inflation is realized, as long as the height of the potential, at the time at which density
perturbations are produced, is given by the SM Higgs field.  So, the ingredient (ii) can be changed, affecting only the prediction on the spectral index of density perturbations $n_S$. 
In fact we present here an alternative model within standard Einstein gravity which also provides a graceful exit from Inflation and gives rise to a radiation era, 
without affecting the ingredient (i) and its predictions.

In order to achieve enough Inflation it is necessary to have the Higgs field $\chi$  trapped at a value $\chi_0$ of order $10^{15}-10^{17}$ GeV 
with a suppressed tunneling rate ($\Gamma\ll H^4$, where $\Gamma$ is the tunneling probability per unit time and volume and $H$ 
is the Hubble rate) and then it is necessary, after at least 60 e-folds, to trigger a phase transition through an additional scalar which plays 
the role of a clock, so that $\chi$ is not stuck at $\chi_0$ anymore, either exiting through tunneling or through classical roll. When this 
happens the field $\chi$ can roll down fast to smaller values, ending thus Inflation, and eventually oscillate around zero and dissipate 
energy, producing thus a hot plasma, and finally relax at its true minimum at the usual field value $\chi=246$ GeV.

There are two ways to get such a behavior: (a) the tunneling rate $\Gamma$ is constant and smaller than $H$ during Inflation and then $H$ has to decrease until at some time 
$\Gamma \sim H^4$, so that an efficient phase transition can occur through bubble nucleation; this is what we have proposed 
in~\cite{Masina:2011aa}, based on~\cite{oldinfl1,oldinfl2}, using a Brans-Dicke scalar $\phi$ which evolves and slows down the expansion; 
(b) $\Gamma$ is time-dependent while the Hubble rate stays roughly constant, which can be realized in usual Einstein gravity; here the Higgs field interacts directly with some additional degree of freedom $\Phi$; such a field has a 
time evolution and modifies slightly the shape of the barrier so that at some specific value $\Phi_F$ the Higgs can tunnel or even
just classically roll down.

In the present paper we introduce the possibility (b), similarly to what happens in the so-called hybrid inflation scenarios~\cite{LIN2SC,Lyth:1998xn}, where a  ``waterfall'' field is trapped at zero value and can start evolve suddenly when another field (in this case $\Phi$) reaches some trigger value $\Phi_F$. The difference here is that we consider the trapped field to be the Standard Model Higgs field, which is stuck  at a very large value $\chi_0$, instead of being trapped in zero.

Such a possibility could be viewed as less minimal than the one proposed previously in~\cite{Masina:2011aa} because it introduces an explicit coupling between  the Higgs field and the extra field $\Phi$, which could potentially alter significantly the Standard Model and in particular it could change the RGE themselves and so the existence or location of the False Vacuum. However, as we will see, the coupling needs to be very weak and the mass of $\Phi$ very large, so the contribution to the RGE is practically zero, leaving thus the connection between low-energy parameters ($m_H$ and $m_{t}$) and the false vacuum unchanged.
Moreover while the case analyzed in~\cite{Masina:2011aa} has the virtue of having only gravitational couplings and of having a rather sharp prediction on the spectral index of density perturbations $n_S$, nonetheless it has the complication that the additional Brans-Dicke field goes to very large values and it is necessary to introduce some additional explicit potential, in order to make $\phi$ roll down again towards zero value after Inflation, so that fifth-force constraints due to an extra light scalar can be satisfied. 
This introduces some model-dependence in the post-inflationary evolution and features an uncertainty on the number of e-folds needed.
In the case we present here instead there is no problem with the late-time behavior of $\Phi$, because gravity is standard.
Moreover it is also worth to note that hybrid inflation has the feature that it can avoid the fields to have superplanckian excursions 
during Inflation, which is usually what happens in simple single-field slow-roll models, and also what happened in our 
model \cite{Masina:2011aa} for the field $\phi$.
While we share the point of view that this is not necessarily a problem, because in any case the energy densities are always smaller 
than Planckian (see {\it e.g.}~\cite{Lindebook}, sect.2.4 for a discussion), we note the interesting fact that we can have two regimes, 
depending on the Higgs  mass.
As we are going to show in fact, we can get Inflation to work with subplanckian field excursion only for low values of $m_H$.

The paper is organized as follows. 
In section~\ref{Model} we present the inflationary model, with the Higgs field and an extra field $\Phi$. 
We impose the constraints that the $\Phi$ field has to slowly roll down its potential for  enough e-folds and has to generate the correct 
amplitude of density perturbations.  We then compute the spectral index $n_S$ as a function of the parameters and show 
the regions of parameter space consistent with all observations for various values of $m_H-m_t$.
In section \ref{classical} we show that introducing a small coupling between the potential barrier along the Higgs field direction can be erased efficiently.
We finally draw our conclusions in section \ref{conclusions}.

\section{Higgs hybrid model}
\label{Model}
\medskip{}

We consider here the usual Standard Model Higgs field potential 
\be 
V_{\rm Higgs}(\chi)= \frac{\lambda}{24}(\chi^{2}-v_\chi^{2})^{2} \,\,\,, 
\ee
including quantum corrections due to the running of its quartic coupling $\lambda$, as in~\cite{CERN-TH-2683,hep-ph/0104016}.
It is known that for a narrow range in parameter space $m_H-m_t$ there is a new local minimum at a value $\chi_0$, 
at very high energies~\cite{strumia2,Masina:2011aa,Masina:2011un}. 
Such a range is realized along a line in parameter space 
\be
m_t [{\rm GeV}]=111.835  + 0.479\,\, m_H [{\rm GeV}] \,\,
\ee
where an error of about 3 GeV on $m_H$ is present due to theoretical uncertainties 
on the RGE. 
We assume that the Higgs field $\chi$ starts trapped in a cold coherent state in this false minimum, so that the Universe is dominated 
by the potential energy $V_{\rm Higgs}(\chi_0)$ and thus can inflate. 
It is however necessary to end Inflation and have a transition 
to a radiation dominated era and in order to do that we introduce here a new scalar field $\Phi$ which is very weakly coupled 
to the Standard Model Higgs.

Such additional field $\Phi$ evolves with time and makes the barrier in  $V_{\rm Higgs}(\chi)$ disappear after at least $60$ e-folds. 
This can be achieved if the evolution of $\Phi$ is sufficiently slow and it is similar to what happens in the hybrid inflation scenarios, 
although usually the ``waterfall'' field is trapped at zero value, while here we consider a very large value $\chi_0$. 
In fact we assume that $\Phi$ is coupled to the Higgs field in such a way that, when it reaches a value $\Phi_F$, the false minimum 
is erased and the Higgs can start rolling down its potential.
The simplest interaction term that we can introduce among the two fields is:
$
V_{\rm int}=\frac{\alpha}{2}\Phi^{2}\chi^{2} \,\,,
$
where $\alpha$ is some dimensionless parameter.
In order to make the Higgs potential  steeper and so erase the barrier, we have two options: (i) $\alpha$ positive and $\Phi$ growing from zero to large values or (ii) $\alpha$ negative and $\Phi$ decreasing from large to zero value. 

We study here the first option, which would be analogous to so-called inverted hybrid inflation scenario~\cite{Lyth:1996kt}. 
Note that the interaction term $V_{\rm int}$ acts as a positive mass for $\Phi$, while we would want the field to be unstable. 
For this reason we introduce self-interactions $V_\Phi$ so that $\Phi$ can have a negative squared mass and so start close 
to zero and then increase:
\begin{equation}
 V_{\Phi}(\Phi)=  \frac{\sigma}{24}(\Phi^{2}-v_\Phi^{2})^{2}\,\,.
 \label{eq:Vphi}
\end{equation}
Hence, for the full scalar potential we would get
\begin{equation}
V(\Phi,\chi)=V_{\rm Higgs}(\chi)+V_{\rm int} + V_{\Phi}(\Phi)= \frac{\lambda}{24}(\chi^{2}-v_\chi^{2})^{2}+ \frac{\alpha}{2}(\Phi^{2}-v^2_\Phi)\chi^{2} 
+  \frac{\sigma}{24}(\Phi^{2}-v_\Phi^{2})^{2}\,\,,
 \label{eq:vtotal0}
\end{equation}
where we have also shifted the interaction term 
\begin{equation}
V_{\rm int}=\frac{\alpha}{2} (\Phi^{2}-v^2_\Phi)\chi^{2} \,\,,
\label{eq:potencialfacil}
\end{equation}
in such a way that its vacuum expectation value vanishes when $\Phi=v_\Phi$ thus recovering the usual Standard Model.

We assume that the  tunneling rate in the $\chi$ direction is very small initially, which can be obtained by tuning  the barrier in $V_{\rm Higgs}$  varying $m_H$ and $m_t$, as we are going to show in the next section. So we can consider $\chi=\chi_{0}$ during Inflation. 
We also assume that initially $\Phi$ starts very close to zero.
We take as the initial value  $\Phi_{in} \simeq H$ which is the minimum value given by quantum fluctuations.
The squared mass term for the field $\Phi$ is simply given by
\begin{equation}
\frac{\partial^2 V(\Phi,\chi)}{ \partial \Phi^2}= \alpha \chi_0^{2} + \frac{\sigma}{3}\Phi^{2}    + \frac{\sigma}{6}(\Phi^{2}-v_\Phi^{2}) \,\,.
\end{equation}
We assume now $H\ll v_{\Phi}$  (we will check later that this is consistent), so that the squared mass initially is just:
\begin{equation}
\left. \frac{\partial^2 V(\Phi,\chi)}{ \partial \Phi^2}\right|_{\Phi=\Phi_{in}} \simeq \alpha \chi_0^{2} - \frac{\sigma}{6} v_\Phi^{2}\,\,,
\end{equation}
so that the potential for the field $\Phi$ is tachyonic initially provided that 
\begin{equation}
\alpha < \frac{\sigma}{6}   \left(\frac{v_{\Phi}} { \chi_0}\right)^2  \, .
\label{boundalpha}
\end{equation}
We can also rewrite the potential when $\chi=\chi_0$ as:
\begin{equation}
V(\Phi,\chi_0)=V_{\rm Higgs} (\chi_0)+  \frac{\sigma}{24}(\Phi^{2}-\bar{v}_\Phi^{2})^{2} \label{eq:vtotal} + C\,\,\,,
\,\,\,\bar{v}^2_{\Phi}\equiv v_{\Phi}^2-\frac{6 \alpha \chi_0}{\sigma}\,\,\,, \,\,\,
C = \frac{\alpha \chi^2_0 v_{\Phi}^2}{2} \left( \frac{24}{\sigma}-1 \right)-\frac{36 \alpha^2 \chi^4_0}{\sigma^2}\,\,.
\end{equation}

We also assume that  $\frac{\sigma \bar v^4_{\Phi}}{24} \ll V_{\rm Higgs}(\chi_0)$ so that the energy density stored in $V_{\rm Higgs}$ is much 
larger than the one in $V_{\Phi}$ , otherwise we would simply have a single-field slow-roll model in the $\Phi$ direction, which does not use the existence of the false minimum in the Higgs potential and which can work only for superplanckian field values. If the Higgs dominates, then the Hubble rate $H_I$ during Inflation is practically exactly constant and given by:
\be
H_I^2\simeq \frac{V_{\rm Higgs}(\chi_0)}{3 M^2} \,\, ,
\ee
where $M$ is the reduced Planck mass, $M\approx 2.43 \times 10^{18}$ GeV.
In particular the value of the potential is related to the Higgs mass via the approximate relationship:
\begin{equation}
m_H [{\rm GeV}]= 4.867 \,\,{\rm Log}_{10}  \left(  V^{1/4}_{\rm Higgs} (\chi_0) [{\rm GeV}]  \right) + 47.41 \, \, ,
\end{equation}
with the usual theoretical uncertainty of about 3 GeV on $m_H$.

The height of the potential $V_{\rm Higgs}(\chi_0)$ determines thus also the amount of gravitational waves from inflation. 
Using the fact that the amplitude of the scalar density fluctuations is measured, this fixes the value of the tensor-to-scalar ratio 
$r$ to an observed value $r_0$, given as a function of $V_{\rm Higgs}(\chi_0)$ as in~\cite{Masina:2011un} and this leads to a prediction 
on $r$ given by:
\begin{equation}
m_H[{\rm GeV}]= 1.21\,\, {\rm Log}_{10}(r_0) + 127.84 \,\, ,
\end{equation}
again subject to the 3 GeV uncertainty on $m_H$.

Given such initial conditions, $\Phi$ slowly rolls down towards its minimum $\bar{v}_{\Phi}$ according to the equation of motion
\begin{equation}
\ddot{\Phi}+3H\dot{\Phi}+\frac{\partial V}{\partial\Phi}=0\,\,.
\label{eq:dynamics}
\end{equation}

\begin{figure}[h!]
 \includegraphics[width=6.7cm]{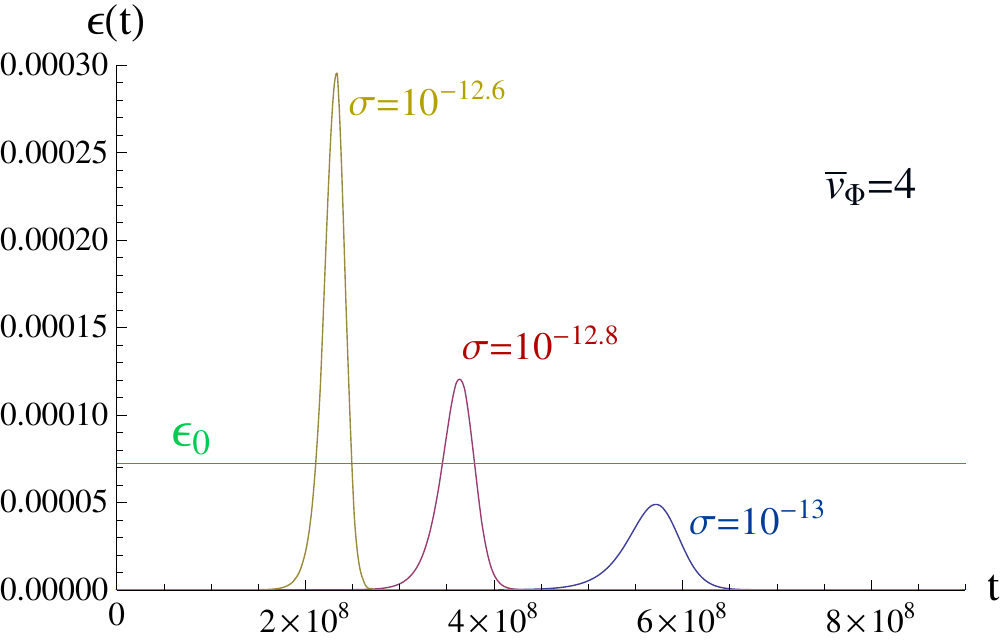} 
\caption{Slow roll parameter $\epsilon$  as a function of time $t$ for $m_H= 124.323915$ GeV, $m_t=171.4$ GeV, ${\bar v}_\Phi=4$ 
and different values of $\sigma$, as indicated by the labels. Here $\bar{v}_{\Phi}$ is expressed in units of 
$M\approx 2.43 \times 10^{18}$ GeV.}
\label{fig-sol}
\vskip .2 cm
\end{figure}

The number of e-folds is computed starting from a given time  $t_0$ is $N\equiv \int_{t_0}^{t} H(\tilde{t}) d\tilde{t} \simeq H_I (t-t_0)$, 
since $H=H_I$ is constant.
During such a stage adiabatic density and tensor fluctuations due to $\Phi$ are produced, according to the the slow-roll parameters, 
given as usual by
\begin{equation} 
\epsilon = \frac{1}{2} \left| \frac{1}{ V}  \frac{d V}{d(\Phi/M) }\right|^2 \,\,\,,\,\,\,
\eta= \frac{1}{ V}  \frac{d^2 V}{d(\Phi/M)^2 }\,\,.
\end{equation}
Using the fact that $r=16 \epsilon$, we know that $\epsilon$ must have a fixed value $\epsilon_0=r_0/16$. 
We can define  $t_0$ as the time at which $\epsilon(t)=\epsilon_0$. 
We show $\epsilon(t)$ in fig.~\ref{fig-sol}.

Note that for each point in parameter space $(\sigma,\bar{v}_{\Phi})$ there can be zero, one or two 
solutions for $t_0$. If there is no solution ($\epsilon$ always smaller than $\epsilon_0$) 
then the mechanism cannot work. If instead $\epsilon(t)$ is large enough there are one or two values of $t_0$.
Given one of them, we assume this corresponds to the observed cosmological scales (precisely to the pivot scale 
$L=1/(0.002)$ Mpc) and we check if  $\bar{N}$ e-folds later (where $\bar{N}$ is the number of e-folds needed to reach the end of Inflation and is close to 60, 
as it is computed in the next section) the field $\Phi$ is still time-dependent at some value\footnote{  We define $\Phi_F$ as 
the field value 60 e-folds (or more exactly the number of e-folds given by eq.~(15)) 
after the time $t_0$, under the condition that $\Phi_F$ is still far from $\bar{v}_{\Phi}$; more precisely, 
we require that $\frac{\bar{v}_{\Phi}-\Phi_F}{\bar{v}_{\Phi}}$ is more than a percent.}
 $\Phi_F<\bar{v}_{\Phi}$. At this point the Higgs field can go out of its false minimum and Inflation ends,
provided that the potential barrier in the Higgs mass is such that it is erased when $\Phi=\Phi_F$. 
We show in the next section that this can be obtained even with a very small $\alpha$, as given by the bound (\ref{boundalpha}).
If  instead the field $\Phi$ has already stabilized at the constant value $\Phi_F \simeq \bar{v}_{\Phi}$ before that $\bar{N}$ e-folds have passed, 
then the mechanism cannot work.

We have checked that at later stages the potential $V(\chi, \Phi>\Phi_F)$ has a first derivative $d V/d\chi$ which is generically 
large (since the typical scale of the problem is $V^{1/4}$, which is much larger than $H$) except in the very fine tuned cases,
 which we have discarded, in which $\frac{\bar{v}_{\Phi}-\Phi_F}{\bar{v}_{\Phi}}$ is very small. 
 We have checked that such a derivative is very large, in the region of parameter space that we have considered, 
 and that the field rolls down in a negligible number of e-folds.

It is possible that the field tunnels along the $\chi$ direction, before rolling down classically. In this case the criterion for the end of Inflation would be to have a tunneling rate per unit time and volume larger than $H^4$. However analyzing in detail such dynamical process is beyond the scope of our paper. 
In fact this happens only at the end of Inflation, which corresponds to very small scales today, and therefore not observable by any means in the spectrum of perturbations. 
The only way this could affect our predictions would be if the tunneling event takes place many e-folds before $\Phi_F$.  
We have checked this using an estimates of the tunneling rate of the type $\Gamma=M^4 E^{-S}$ where $M$ is a typical scale of the problem, 
which we take to  be $V^{1/4}$.  Here we have estimated $S$ using two different mechanisms: the Coleman instanton~\cite{Coleman}, 
where $S$ is the classical bounce action along the $\chi$ direction (note anyway that this is only a qualitative procedure, 
as the problem becomes a two-field tunneling problem and rigorously we should consider both directions in field space when 
finding a classical bounce) and the Hawking-Moss instanton where 
$S=24 \pi^2 M_{Pl}^4 (1/V(\chi_M,\Phi) - 1/V(\chi_m, \Phi))$, where $\chi_M$ and $\chi_m$ 
are the locations of the maximum and the minimum along the $\chi$ direction respectively~\cite{HawkingMoss}. 
Using both estimates we have checked that $\Gamma$ becomes of order $H^4$ only a fraction of e-folds before the classical roll, 
so that our approximations is reliable for our purposes.

\begin{figure}[h!]
 \includegraphics[width=6.7cm]{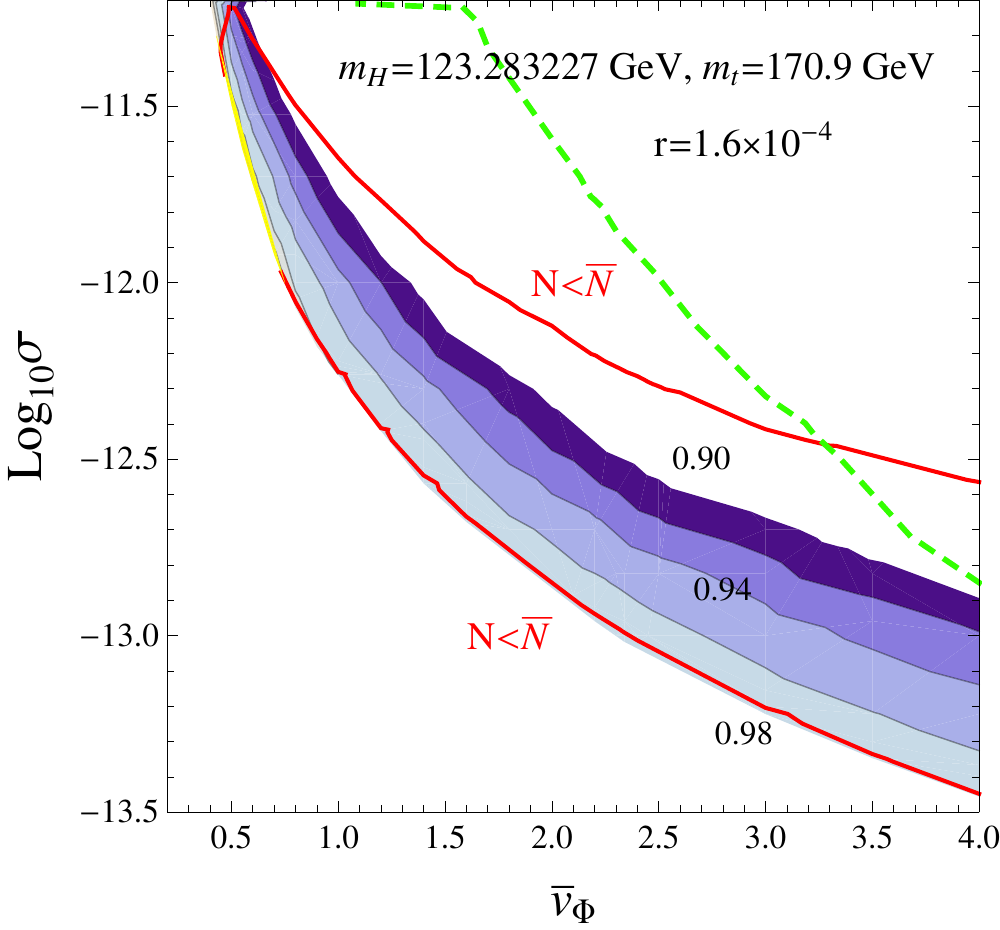} \,\,\,    \includegraphics[width=6.7cm]{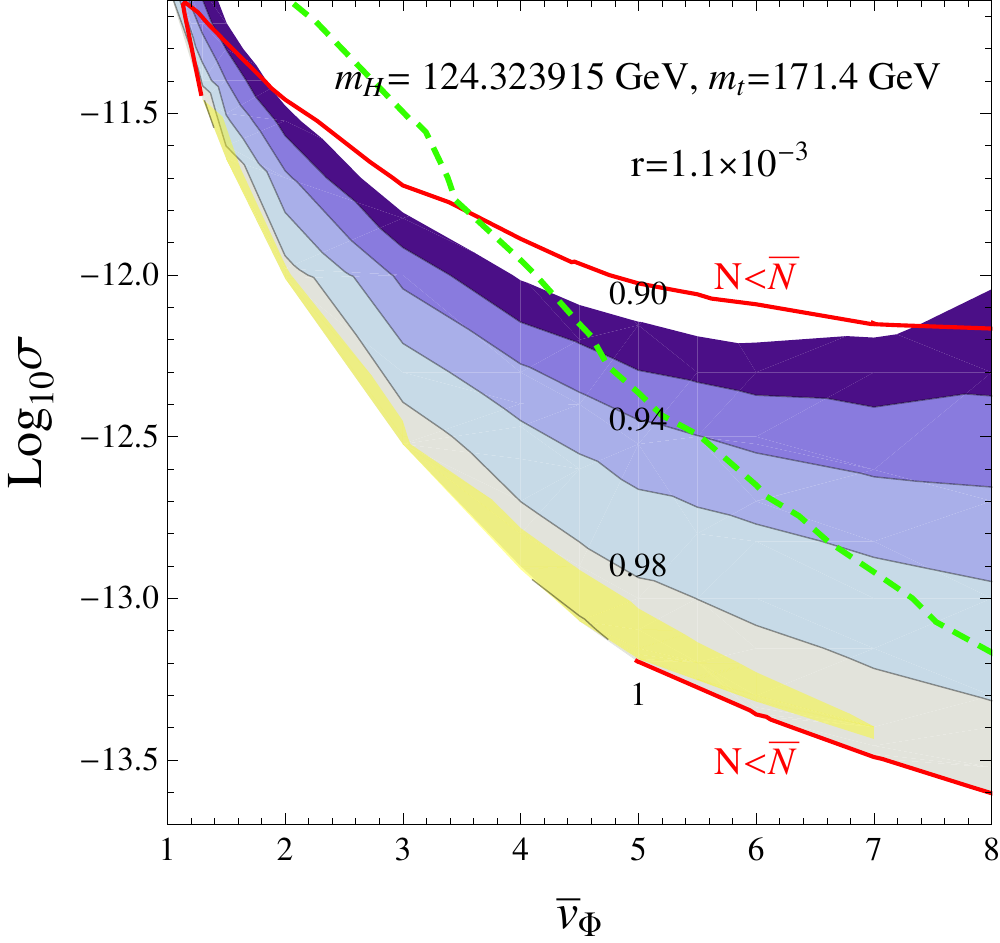} \,\,\, \vskip.3 cm  \includegraphics[width=6.9cm]{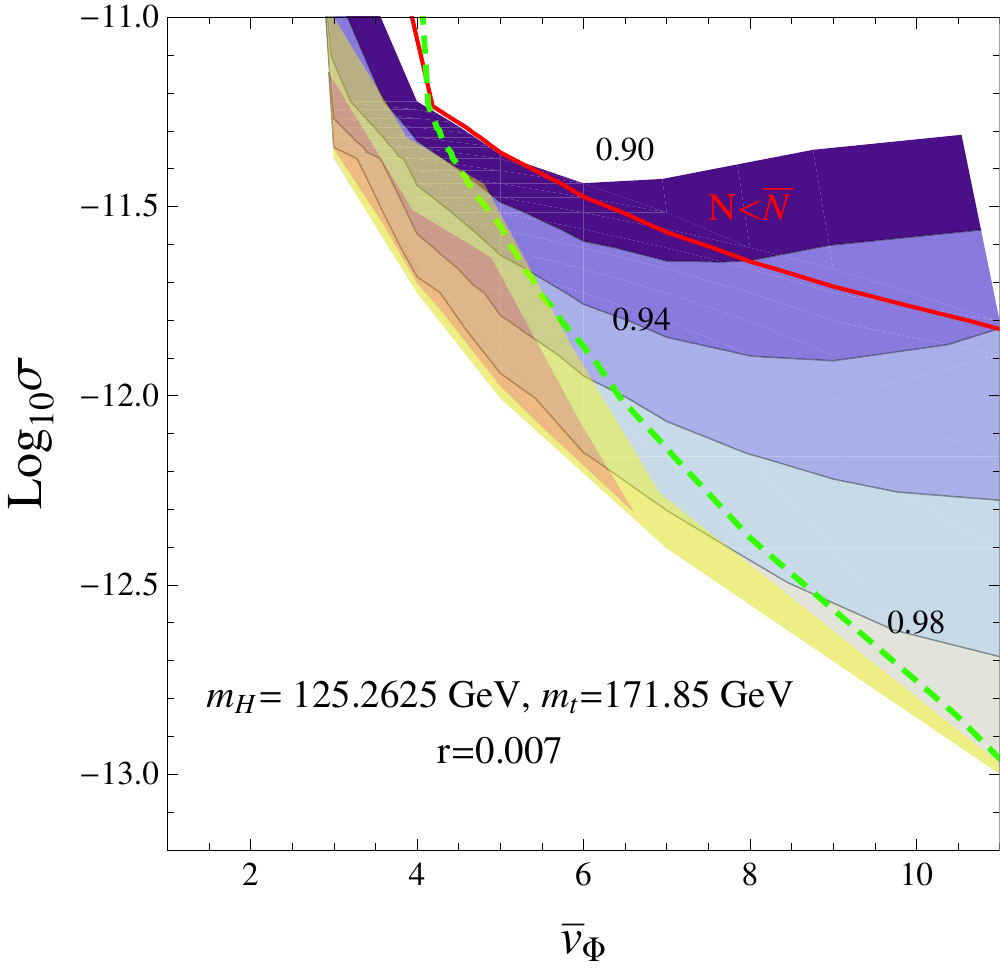} 
\caption{We show here the allowed parameter space for $\sigma, \bar{v}_{\Phi}$, and the corresponding values of the spectral index $n_S$,
for which we display contours ranging from $0.90$ up to $1$ with steps of $0.02$. Here $\bar{v}_{\Phi}$ is expressed in units of 
$M\approx 2.43 \times 10^{18}$ GeV.
The shaded regions (superimposed to the regions where $n_S$ is slightly smaller than one) are such that there is also another solution with $1\le n_S \le 1.02$ (yellow) and  $1.02\le n_S \le 1.04$  (orange).
Each plot has a prediction for $r$ and corresponds to the value shown for $m_H$ and $m_t$, up to the theoretical uncertainties 
of the RGE of about 1 GeV for $m_t$ and $3$ GeV for $m_H$. The regions outside the solid (red) curves where the number of e-folds 
turns out to be smaller than $\bar N$ are excluded. The region at the right of the dashed (green) curve is also excluded since there 
the contribution to potential $V(\Phi,\chi_0)$ is not dominated by $V_{\rm Higgs}(\chi_0)$; for the sake of definiteness we have drawn the 
dashed line where $\sigma \bar v^4_{\Phi}/{24}  \approx V_{\rm Higgs}(\chi_0) /3$. Note also that the values of $\bar{N}$ in the figure have been calculate assuming that the field $\Phi$ decays rapidly and never dominates after Inflation, as discussed in section \ref{classical}; other scenarios of decay would shift such number by a few, depending on the energy density stored in $\Phi$ at the final value $\Phi_0$, so the shift would be minimal for small $\bar v_{\Phi}$ and close to the solid (red) curve, where $\Phi_0$ is close to $\bar v_{\Phi}$. }
\label{fig-ns}
\vskip .2 cm
\end{figure}

When there are solutions for $t_0$ we can also compute the spectral index $n_S=1+2 \eta(t_0)-6 \epsilon(t_0)$ and check if it is consistent with observations, for the two solutions. The results are shown in fig.~\ref{fig-ns}. The first solution gives rise to $n_S<1$ and results consistent with observations are indeed found, since the experimentally 
preferred region for $n_S$ is in the window $0.93-1$. The other solution gives instead an $n_S$ slightly bigger than 1, as emphasized by means of the (yellow and orange) shaded regions.
The regions outside the solid (red) curves are excluded because the number of e-folds turns out to be smaller than $\bar N$. 
Also the region at the right of the dashed (green) curve is not of interest, since here the contribution to potential $V(\Phi,\chi_0)$ 
is not dominated by $V_{\rm Higgs}(\chi_0)$ but rather by $V_\Phi(\Phi)$, see eqs.  (\ref{eq:vtotal0}) and (\ref{eq:vtotal}). 
Notice that, by increasing $m_H$ and $m_t$, the region in the $\sigma-\bar v_\Phi$ plane consistent with the observed window for $n_S$
is shifted to the right, namely to higher values of  $\bar v_\Phi$. But then it becomes more difficult to satisfy the constraint 
$\sigma \bar v^4_{\Phi}/{24}  \ll V_{\rm Higgs}(\chi_0)$.  For the sake of definiteness we have drawn the 
dashed (green) curve where $\sigma \bar v^4_{\Phi}/{24}  \approx  V_{\rm Higgs}(\chi_0)/3$, which cannot be satisfied  
for values of $m_H \gtrsim 125.3$. 

An interesting point is that if $m_H$ and $m_t$ are large so also $V_{\rm Higgs}(\chi_0)$ and $r$ are large and in this case 
it turns out that the values needed for $\bar{v}_{\Phi}$ have to be larger than the reduced Planck mass, $M$. 
While this is not necessarily a problem, because in any case the energy densities are always smaller than Planckian, we note the interesting fact that, imposing this situation to be avoided, would select an upper bound on $r$ of about $0.001$ and on the Higgs mass of about $m_H=124$ GeV   as shown in fig. \ref{fig-ns}. 
Remind however that an upper bound on the Higgs mass cannot be taken at face value because of the theoretical uncertainties on the RGE, 
so a conservative upper bound would be of about $127$ GeV with the present theoretical uncertainties.

Finally note that the above results could be subject to some uncertainty on the precise calculation of $\bar{N}$. 
As we are going to explain in the next section in fact we must know the entire post-inflationary evolution to have the precise value for $\bar{N}$, 
but this depends on the couplings of the field $\Phi$ (if it can decay to some other species through direct couplings or only gravitationally). The results obtained in fig.~\ref{fig-ns} are valid in the case that $\Phi$ decays immediately after Inflation and never dominates the Universe, while otherwise they may receive some correction, as we discuss below.

\section{End of Inflation and Post-inflationary evolution}
\label{classical}

\begin{figure}[th!]
\includegraphics[width=7cm]{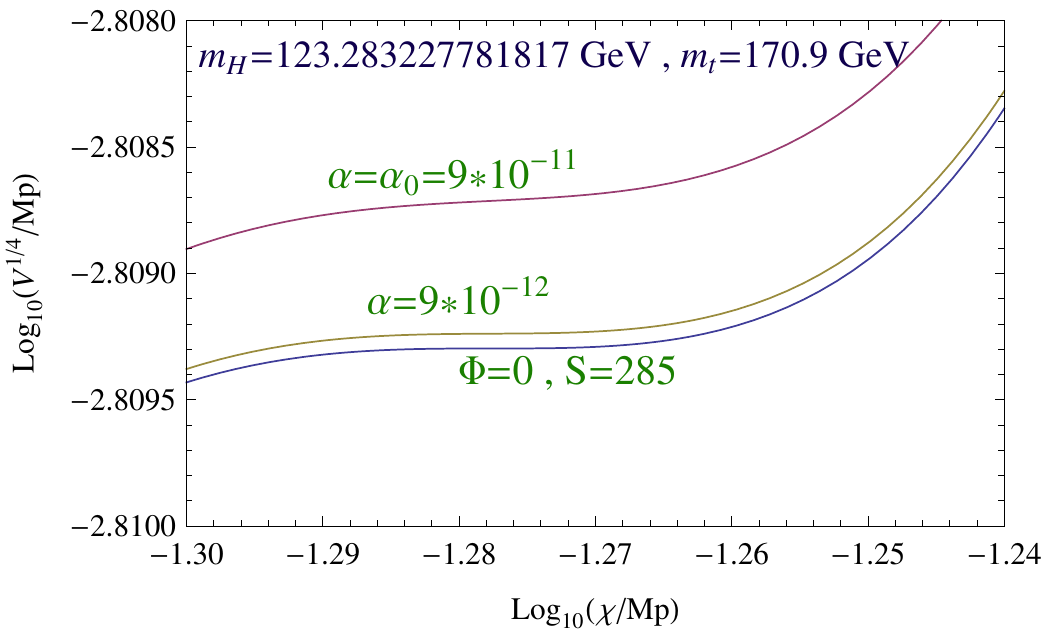} \,\,\,\,\, \,\,
\includegraphics[width=7cm]{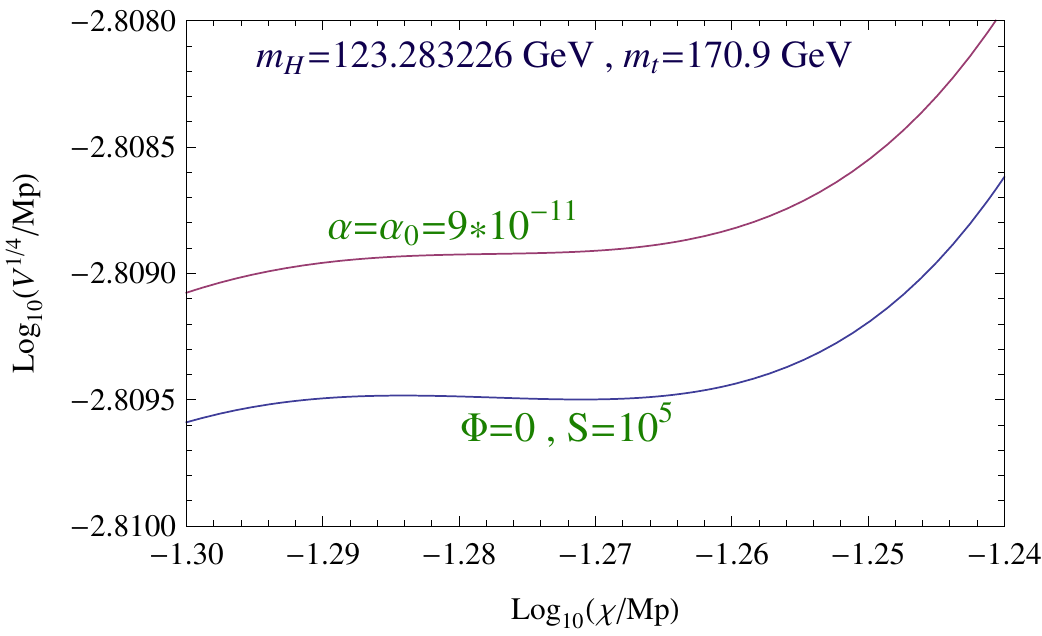}  \\  \vskip .4cm
\includegraphics[width=7cm]{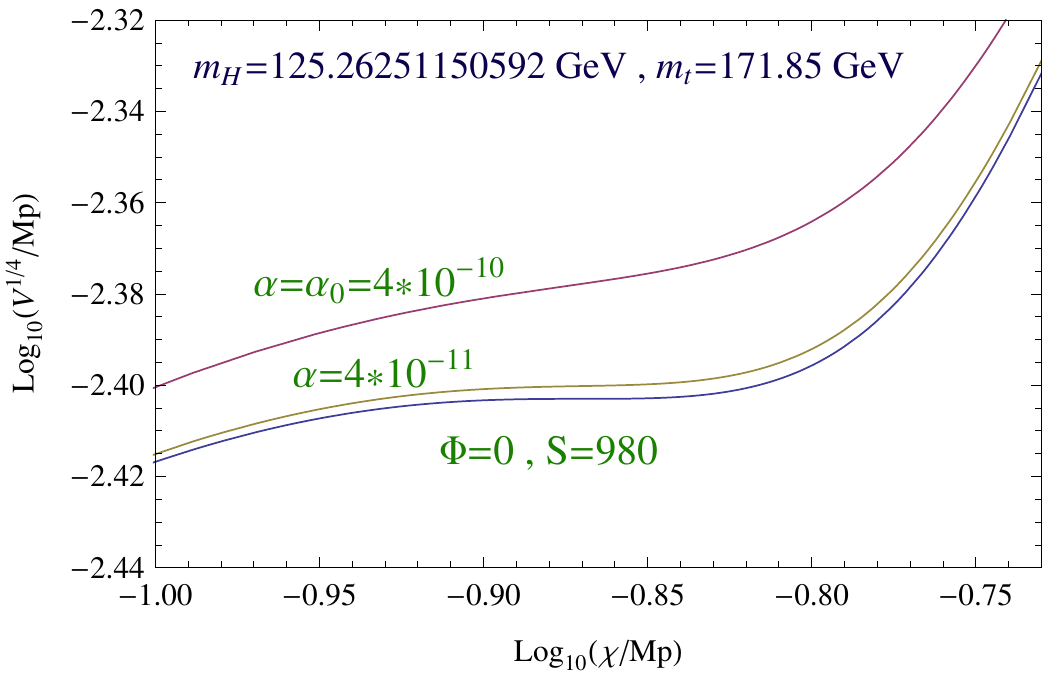}  \,\,\,\,\, \,\,
\includegraphics[width=7cm]{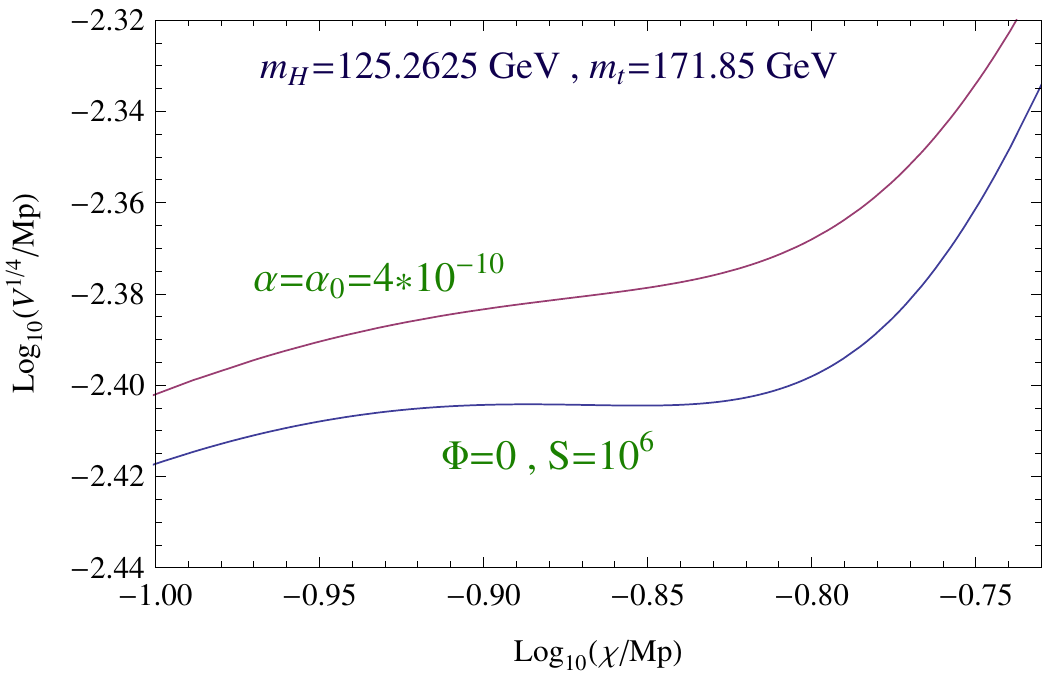}
\caption{ We show the potential $V(0,\chi)$ (bottom blue curves) for
which the field is trapped in the $\chi$ direction in a minimum by a
barrier and the potential at the end of Inflation
$V(\bar{v}_{\Phi},\chi)$, which has no barrier anymore, for several
values of the coupling $\alpha$. The top purple lines correspond to
the maximal possible value for $\alpha$, which we call $\alpha_0$, as
given by eq.~(\ref{boundalpha}). The left top panel represents a case
with a very fine-tuned shallow barrier: here the mass of the field in
the minimum is of about  $1.5 \times 10^{13} GeV$, while $H = 1.38
\times 10^{13} $ GeV and the tunneling rate computed with a Coleman
instanton~\cite{Coleman} has an exponential suppression $\Gamma \propto
e^{-S}$, with an action $S=285$; in this case the minimal $\alpha$
needed to lift the barrier is of about $10^{-5} \alpha_0$.  The left
bottom panel is the analoguos case with a higher $m_H$: here the mass
of the field in the minimum is of about  $7 \times 10^{13} GeV$, while
$H = 9 \times 10^{13} $ GeV and the action is $S=950$.
The right top panel represents instead the potential with maximum
depth, which can be lifted by an $\alpha=\alpha_0$. Here the mass of
the field in the minimum is of about  $4.5 \times 10^{14} GeV$, while
$H = 1.38 \times 10^{13} $ GeV and the tunneling rate has a much
stonger exponential suppression with $S\simeq 10^5$.
The right bottom panel is  also a potential mith maximal depth, with
higher $m_H$: here the mass of the field in the minimum is of about
$2 \times 10^{15} GeV$, while $H = 9 \times 10^{13} $ GeV and the
action is $S\simeq 10^6.$}
\label{fig-alfa}
\end{figure}

Given a value of the height of the potential $V_{\rm Higgs}(\chi_0)$, of $\sigma$ and $ \bar{v}_{\Phi}$ and given a value of $\alpha$ which respects the bound (\ref{boundalpha}) we can finely tune $m_H$ as compared to $m_{t}$, in order to obtain a potential $V(\Phi,\chi)$ such that the barrier in the $\chi$ direction disappears exactly at $\Phi=\Phi_F$ ending thus Inflation. Moreover the barrier at the beginning has to be high enough so to have $\Gamma\ll H_I^4$.
It turns out that this can be done, because even if $\alpha$ is very small and so the change in $V$ is tiny (so that also $V_{\rm Higgs}(\chi_0)$ can be considered constant during Inflation), such change is sufficient to go from a situation where $\Gamma$ is extremely suppressed and practically zero to the situation where the barrier even disappears. This happens because the dependence of $\Gamma$ on the barrier is exponential. Note that there is an intermediate stage when the barrier is still present, but $\Gamma$ is larger than $H_I^4$ so that the transition could start through nucleation of bubbles. However such a situation is extremely short in time and the bubbles would be nucleated at very small scales with no observational interest, so it can be ignored for practical purposes.
We show a typical behavior of the potential in fig.~\ref{fig-alfa} for several values of the parameters.
Only in the case in which $\alpha$ is much smaller than the bound (\ref{boundalpha}),  the shift in the barrier can be insufficient.

Let us discuss what happens after Inflation ends. 
Since the Higgs potential is steep, we can assume that when the barrier disappears the Higgs field rapidly rolls down, dissipates energy and relaxes to $v_\chi=246$ GeV. 
We assume therefore an instantaneous transition from exponential inflation to a radiation dominated phase with scale factor $a(t)\propto t^{1/2}$. 
Note however that when the Higgs starts to roll down classically also the $\Phi$ field undergoes further evolution going from the value $\Phi_F$ to the minimum $v_{\Phi}$. 
The evolution can be solved assuming that the energy density of the Higgs has been transformed into radiation, showing that the field $\Phi$ undergoes oscillations about the minimum and therefore redshifts as matter. For this reason, at some point the field $\Phi$ can also become dominant, although this depends on the amplitude of the oscillations, which is set by the initial value $\Phi_F$. As the Universe expands the field $\Phi$ would dominate the Universe at some point, unless it decays before into relativistic degrees of freedom. The field $\Phi$ is coupled to the Higgs field via eq.~(\ref{eq:potencialfacil}) and so it could decay into quanta of the Higgs field, which are relativistic at high temperature when the Higgs has relaxed in its minimum. The decay rate of the $\Phi$ field could happen at least perturbatively by decaying into two quanta of the Higgs field with a rate $\Gamma_{\chi \chi} \propto  \alpha^2 \mu$, where $\mu$ is some mass scale, but it could perhaps also decay more rapidly due to non-perturbative parametric resonance; moreover we could also assume the field $\Phi$ to have additional couplings which would make it decay rapidly. Finally there is at least also the possibility of decaying through gravitational couplings~\cite{Lindebook} which have a rate $\Gamma_{\rm grav}\approx m_{\Phi}^3/M^2$, where $m^2_{\Phi}=\left. \frac{\partial^2 V(\Phi,\chi)}{ \partial \Phi^2}\right|_{\Phi=v_{\Phi}} \simeq  \frac{\sigma}{3} v_\Phi^{2}\,\,
$. In any case it is easy to see that the field $\Phi$ decays always before nucleosynthesis, at least through gravitational couplings and this happens when $H=\Gamma_{\rm grav}$, which is realized at a temperature of about $T_{\rm decay}\simeq \frac{\sigma^{3/4} v^{3/2}_{\Phi}}{g_*^{1/4} M^{1/2}}$, always higher than about $10^7-10^8$ GeV. It is however possible that the field decays much earlier by one of the above mentioned ways; we leave such study for a more detailed future analysis.

Let us finally compute the number of e-folds $\bar{N}$ needed to go from the time $t_0$ (at the pivot scale taken to be  $L=1/(0.002)$ Mpc) to the end of Inflation, when the barrier for the Higgs field disappears and the field can start rolling down its potential. At this point there is some model dependence which can shift $\bar{N}$ by a few, due to the fact that the field $\Phi$ can dominate and then decay, as discussed above. If we assume that $\Phi$ can rapidly decay and never dominates we can easily compute $\bar{N}$: this is the regime which we have assumed in order to find the results shown in fig.~\ref{fig-ns}.
Let us then compute the time when a particular comoving scale $L$ went outside the horizon during Inflation. In general a scale $L$ leaves
the horizon at some e-folding number $\bar{N}$ if:
\be
L \left(   \frac{T_0}{T_{\rm{RH}}} \right)  e^{- \bar{N}}=\bar{H}_{I}^{-1} \, ,
\ee
where the reheating temperature is simply given by $g_{*} T_{RH}^4  \simeq V_{\rm Higgs}(\chi_0)$, where $g_* = 106.75$ 
is the Standard Model number of degrees of freedom.
This leads to a number of about 60, depending on $V_{\rm Higgs}(\chi_0)$, which is determined by $m_H$.
\\
\\
\\

\section{Conclusions}
\label{conclusions}

The main conclusions of the paper are as follows: (i) it is possible to get Inflation from the Higgs field false vacuum energy density at 
${\cal O}(10^{15} {\rm GeV})^4$, provided a second subdominant field $\Phi$ with a mexican hat potential is (very weakly) coupled to the Higgs field: the field $\Phi$ rolls down its potential and when it reaches a value $\Phi_F$ the barrier in the Higgs potential is erased (ii) this is possible only if the energy scale is not too large and it leads to a bound on the tensor-to-scalar ratio given by $r <0.007$ 
(iii) such a window corresponds to an upper bound on the Higgs mass $m_H<125.3 \pm 3_{th}$ GeV where the $3_{th}$ is the present theoretical uncertainty on the Standard Model RGEs, which together with a lower bound coming from the experimental measurements of the top mass leads to an allowed window $123 \lesssim m_H \lesssim 128$, which could be narrowed down to an interval of 2 GeV by a better theoretical control of the RGE. Such a window corresponds to a range in the  tensor-to-scalar ratio $10^{-4} \lesssim r < 0.007$   (iii) varying the parameters of the $\Phi$ potential, values of the spectral index $n_S$ compatible with the observations can be obtained (iv) the $\Phi$ field excursion is subplanckian only if the Higgs mass is lower than about $124\pm 3_{th}$ GeV, which corresponds to $r<0.001$.


\vskip 1cm
{\bf Acknowledgements} We thank Jaume Garriga  for useful discussions. Work supported by MEC FPA2010-20807-C02-02.



\begin{thebibliography}{15}






\bibitem{Masina:2011aa} 
  I.~Masina and A.~Notari,
  arXiv:1112.2659 [hep-ph]. 

\bibitem{Masina:2011un} 
  I.~Masina and A.~Notari,
  arXiv:1112.5430 [hep-ph]. To appear in Phys.Rev.Lett.
  
  
\bibitem{CERN-TH-2683}
  N.~Cabibbo, L.~Maiani, G.~Parisi and R.~Petronzio,
  Nucl.\ Phys.\ B\ {\bf 158} (1979) 295.
  P.~Q.~Hung,
  Phys.\ Rev.\ Lett.\ \ {\bf 42} (1979) 873.
  M.~Lindner,
  Z.\ Phys.\ C\ {\bf 31} (1986) 295.
  M.~Lindner, M.~Sher and H.~W.~Zaglauer,
  Phys.\ Lett.\ B\ {\bf 228} (1989) 139.
  M.~Sher,
  Phys.\ Rept.\ \ {\bf 179} (1989) 273.
B.~Schrempp and M.~Wimmer,
  Prog.\ Part.\ Nucl.\ Phys.\ \ {\bf 37} (1996) 1
  [hep-ph/9606386].

\bibitem{hep-ph/0104016}
J.~A.~Casas, J.~R.~Espinosa and M.~Quiros,
  Phys.\ Lett.\ B {\bf 382} (1996) 374
  [hep-ph/9603227].
  G.~Isidori, G.~Ridolfi and A.~Strumia,
  Nucl.\ Phys.\ B\ {\bf 609} (2001) 387
  [hep-ph/0104016].
  J.~R.~Espinosa, G.~F.~Giudice and A.~Riotto,
  JCAP\ {\bf 0805} (2008) 002
  [arXiv:0710.2484 [hep-ph]].
     N.~Arkani-Hamed, S.~Dubovsky, L.~Senatore and G.~Villadoro,
  JHEP\ {\bf 0803} (2008) 075
  [arXiv:0801.2399 [hep-ph]].
  
\bibitem{strumia2}
 G.~Isidori, V.~S.~Rychkov, A.~Strumia and N.~Tetradis,
  Phys.\ Rev.\ D\ {\bf 77} (2008) 025034
  [arXiv:0712.0242 [hep-ph]].



\bibitem{oldinfl1}
  F.~Di Marco and A.~Notari,
  Phys.\ Rev.\ D\ {\bf 73} (2006) 063514
  [astro-ph/0511396].

\bibitem{oldinfl2}
  T.~Biswas and A.~Notari,
  Phys.\ Rev.\ D\ {\bf 74} (2006) 043508
  [hep-ph/0511207].


  




\bibitem{LIN2SC} A. D. Linde, Phys. Lett. {\bf B259}, 
38 (1991) . 

\bibitem{Lyth:1998xn} 
  D.~H.~Lyth and A.~Riotto,
  Phys.\ Rept.\  {\bf 314}, 1 (1999)
  [hep-ph/9807278].
  
  \bibitem{Lindebook}
A. D. Linde, ``Particle Physics and Inflationary Cosmology'', Chur, Switzerland: Harwood (1990) 362 p. (Contemporary concepts in physics, 5), arXiv:hep- th/0503203.


  
\bibitem{Lyth:1996kt} 
  D.~H.~Lyth and E.~D.~Stewart,
  Phys.\ Rev.\ D {\bf 54}, 7186 (1996)
  [hep-ph/9606412].
  

\bibitem{Coleman}
   S.~R.~Coleman and F.~De Luccia,
  Phys.\ Rev.\ D {\bf 21}, 3305 (1980).

\bibitem{HawkingMoss}
S.~W.~Hawking and I.~G.~Moss,
  Phys.\ Lett.\ B {\bf 110}, 35 (1982).
  
  
\end{thebibliography}
\end{document}